\documentclass[10pt, conference, letterpaper]{IEEEtran}
\IEEEoverridecommandlockouts

\usepackage{cite}
\usepackage{amsmath,amssymb,amsfonts}
\usepackage{algorithmic}
\usepackage{graphicx}
\usepackage{textcomp}
\usepackage{xcolor}
\usepackage{float}
\usepackage{url}
\usepackage{booktabs}
\usepackage{array}
\usepackage{multirow} 
\usepackage{comment}
\usepackage[caption=false,font=footnotesize]{subfig}

\def\BibTeX{{\rm B\kern-.05em{\sc i\kern-.025em b}\kern-.08em
    T\kern-.1667em\lower.7ex\hbox{E}\kern-.125emX}}

\begin{document}

\title{Blockchain-Driven Federation for Distributed Edge Systems: Design and Experimental Validation}

\author{
\IEEEauthorblockN{Adam Zahir\IEEEauthorrefmark{1}, Milan Groshev\IEEEauthorrefmark{2}, Carlos J. Bernardos\IEEEauthorrefmark{1}, and Antonio de la Oliva\IEEEauthorrefmark{1}}
\IEEEauthorblockA{
\IEEEauthorrefmark{1}Universidad Carlos III de Madrid, Madrid, Spain\\
\IEEEauthorrefmark{2}IE University, Madrid, Spain\\
}}

\maketitle

\begin{abstract}
\emph{Edge computing} brings computation near end users, enabling the provisioning of novel use cases. To satisfy end-user requirements, the concept of \emph{edge federation} has recently emerged as a key mechanism for dynamic resources and services sharing across edge systems managed by different administrative domains. However, existing federation solutions often rely on pre-established agreements and face significant limitations, including operational complexity, delays caused by manual operations, high overhead costs, and dependence on trusted third parties. In this context, Distributed Ledger Technologies (DLTs) such as \emph{blockchain} can create dynamic federation agreements that enable service providers to securely interact and share services without prior trust.

This article first describes the problem of edge federation, using the standardized ETSI \emph{multi-access edge computing (MEC)} framework as a reference architecture, and how it is being addressed. Then, it proposes a novel solution using \emph{blockchain} and \emph{smart contracts} to enable distributed MEC systems to dynamically negotiate and execute federation in a secure, automated, and scalable manner. We validate our framework's feasibility through a performance evaluation using a private Ethereum blockchain, built on the open-source Hyperledger Besu platform. The testbed includes a large number of MEC systems and compares two blockchain consensus algorithms. Experimental results demonstrate that our solution automates the entire federation lifecycle--from negotiation to deployment--with a quantifiable overhead, achieving federation in approximately 18 seconds in a baseline scenario. The framework scales efficiently in concurrent request scenarios, where multiple MEC systems initiate federation requests simultaneously. This approach provides a promising direction for addressing the complexities of dynamic, multi-domain federations across the edge-to-cloud continuum. 
\end{abstract}


\section{Introduction}
\label{introduction}

\emph{Edge computing} introduced a paradigm shift by bringing computation and data processing closer to end users, enabling the provisioning of low-latency, quality-sensitive applications. Its infrastructure spans heterogeneous, geographically distributed resources across multiple administrative domains, including mobile network operators, edge/cloud providers, and application developers. To meet strict real time KPIs, service providers must scale infrastructure dynamically, especially when coverage or demand spikes. In such cases, advanced solutions like \emph{federation} become essential, allowing service providers to extend their capabilities by leveraging external edge systems. Federation is the process of orchestrating services or resources across multiple administrative domains.

In the field of edge computing, \emph{multi-access edge computing (MEC)} is the standardized framework for the deployment and orchestration of edge-native applications. To support federation within the MEC framework, standardization bodies such as ETSI~\cite{etsi-gr-mec-035} and GSMA~\cite{gsma-federation} have proposed a unified reference architecture that introduces a MEC Federator (MEF) component, designed to ensure interoperability of MEC systems across different infrastructure domains. Beyond standardization efforts, several academic works (reviewed in Sec.~\ref{sec:comparison-with-existing-work}) have explored alternative MEF designs and proposed various interfaces for federation~\cite{soa-service-continuity-border-mobility, soa-enhanced-meo, soa-cooperative-service-continuity, soa-mec-federation-resource-sharing}. While their definitions of federation differ slightly, a common assumption is the existence of a pre-established Service Level Agreement (SLA) between service providers. Such agreements typically involve the manual configuration of control and data plane links between MEC systems and the exchange of authentication credentials in advance, which is time-consuming and only suitable for static environments.

In dynamic or large scale settings, providers must establish federation on the fly, without prior trust. In such cases, existing MEF variants suffer from operational complexity, delay due to manual operations, overhead costs, and, in some cases, reliance on trusted third parties. This creates several technical challenges related to security, privacy, availability, billing, and multi-domain QoS enforcement (elaborated in Sec.~\ref{subsec:mec-federation-challenges}). To address these challenges of security, automation, and trust in multi-domain environments, the application of blockchain technology emerges as a promising solution.

Through its inherent transparency and immutability, blockchain can serve as a trustworthy platform for both the negotiation and execution of federation. This paper introduces a blockchain-based MEF designed to support dynamic federation in distributed multi-domain environments. Our design follows a cloud-native approach that removes dependencies on underlying MEC components. We encode federation in a \emph{smart contract (SC)}. Each MEC system runs a node in its MEF and interacts through immutable transactions. This design ensures secure information exchange and supports a fair and efficient competitive process.

The main contributions of this article are:
\begin{itemize}
    \item We define MEC federation as a programmable and secure collaborative agreement implemented through SC, enabling flexible automation and dynamicity in the federation process.
    
    \item We design and integrate a blockchain-based MEF into the ETSI MEC architecture.
    
    \item We validate our proposal through a Proof-of-Concept (PoC) implementation on top of a private Ethereum blockchain and analyze its performance in terms of communication latency, resource consumption, and platform scalability.
\end{itemize}

This article is organized as follows. Sec.~\ref{sec:mec-federation-in-etsi} provides background on MEC federation and outlines its technical challenges in dynamic environments. Sec.~\ref{sec:blockchain-for-mec-federation} explains how blockchain addresses these challenges and presents our architecture design for dynamic and secure MEC federation. Sec.~\ref{sec:experimental-validation} validates the proposal through a PoC implementation and performance analysis. Sec.~\ref{sec:comparison-with-existing-work} compares existing work in the literature. Finally, Sec.~\ref{sec:conclusion} concludes the article.
\section{MEC Federation in ETSI}
\label{sec:mec-federation-in-etsi}

\subsection{MEC in a nutshell}
\label{subsec:mec-in-a-nuthsell}
The MEC framework~\cite{etsi-gs-mec-003} was designed to bring computation closer to end devices, reducing latency and enabling task offloading to a virtualized edge platform with orchestration and service capabilities. Its architecture, illustrated in Fig.~\ref{fig:mec-federation-architecture}, is organized into two main levels: system and host. Hosts can be multiple, and their resources are managed by the system-level components. At the system level, the MEC Orchestrator (MEO) serves as the central authority, maintaining a global view of the MEC hosts, services, and resources. It handles package onboarding and selects the most suitable host for deploying MEC applications. At the host level, the MEC Platform Manager (MEPM) manages the application lifecycle, while the MEP exposes APIs for service discovery and consumption. Underlying compute, storage, and network resources are managed by the Virtual Infrastructure Manager (VIM).

\subsection{MEC federation}
\label{subsec:mec-federation-overview}
\begin{figure}
    \centering
    \includegraphics[width=1\columnwidth]{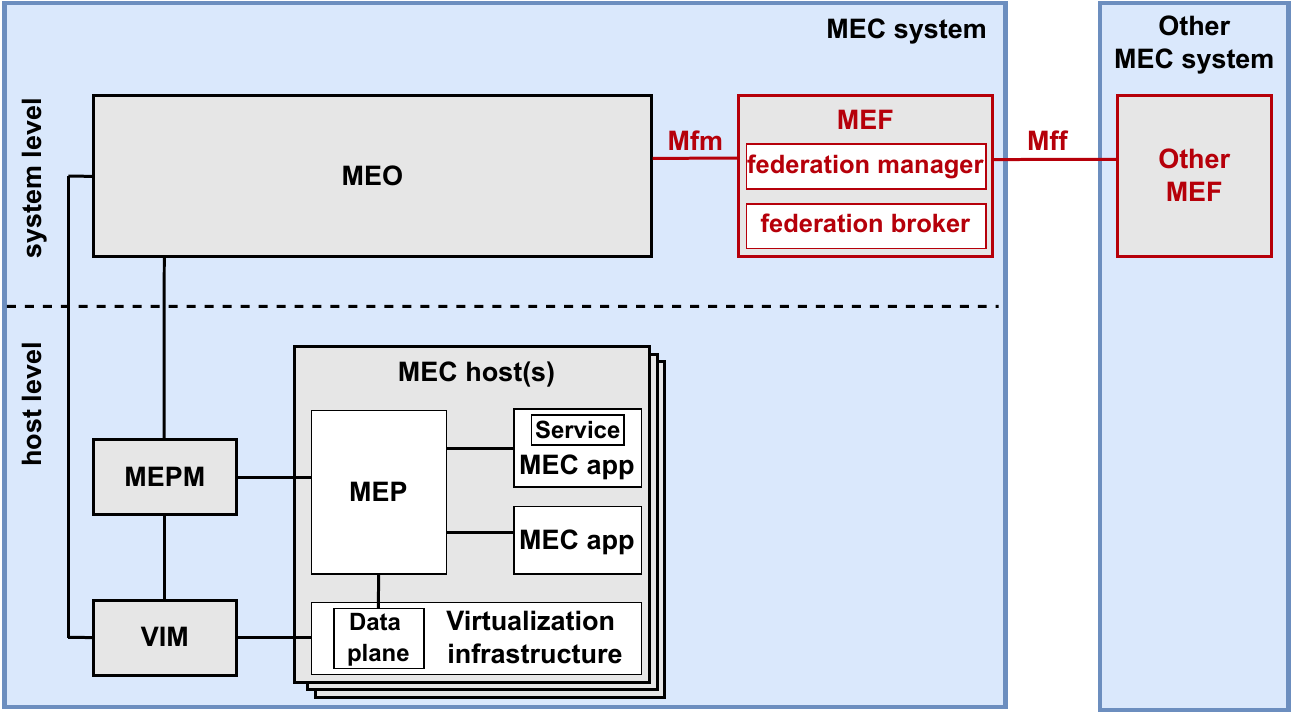}
    \caption{Simplified MEC reference architecture variant for federation}
    \label{fig:mec-federation-architecture}
    \vspace{-1em}
\end{figure}

The federation variant (highlighted in red in Fig.~\ref{fig:mec-federation-architecture}) introduces a new functional element: the MEF, responsible for integrating MEC systems into the federation framework. Each MEF interfaces with at least one MEO via the \emph{Mfm} reference point and coordinates information exchange with other MEFs through the \emph{Mff} reference point. An MEF may assume two roles: a federation manager, which is responsible for authorization of federation members, security, information exchange (e.g., system details), resource and platform discovery, service catalog exposure, and assurance functions (e.g., charging, monitoring); and (optionally) a federation broker, which enables one-to-many interactions between MEFs.

MEC federation coordinates MEC systems across multiple owners through agreements. One core aspect influencing how these agreements are formed is the dynamic nature of the environment. We can identify two main approaches: \emph{pre-established} federation, where MEC providers negotiate agreements and SLAs in advance (e.g., long-term roaming contracts); and \emph{open} federation, where domains can join or leave dynamically without prior negotiation, enabling rapid SLA definition and flexible collaboration (e.g., temporary edge resource sharing to accommodate sudden service demand in specific locations). While most existing solutions in the literature (discussed in Sec.~\ref{sec:comparison-with-existing-work}) focus on pre-established and static federation models based on long-term agreements and trusted relationships, this work addresses the less-explored case of open and dynamic MEC federation.

Once an agreement is reached, the second critical aspect of MEC federation is the control plane interconnection between systems. ETSI has proposed two models for executing federation interactions: centralized and decentralized. In the centralized model, a neutral central entity called the MEC Federation Interconnection Provider (MFIP) serves as an intermediary that manages federation functions (e.g., registration, discovery, lifecycle management, charging). While scalable, this model introduces a costly central entity that poses a single point of failure and requires high trust among all participants. The decentralized model relies on direct peer-to-peer connections between MEC systems. Although simpler, it is time-consuming and lacks scalability, as each new connection requires a business agreement and connectivity setup (e.g., connecting 30 systems may need 30 or more days).

Open federation models require secure and trustworthy frameworks that allow MEC providers to quickly establish dynamic SLAs before service or resource deployment. However, existing centralized and decentralized approaches face several limitations in meeting these requirements (see Sec.~\ref{subsec:mec-federation-challenges}). In this regard, Distributed Ledger Technologies (DLTs) such as \emph{blockchain} offer a promising solution that ensures trust, privacy, and security among users, without relying on a central authority to regulate their interactions.

\subsection{Technical challenges in open and dynamic MEC federation} 
\label{subsec:mec-federation-challenges}
Below, we outline the main challenges posed by MEC federation in dynamic environments and how existing centralized and decentralized interconnection models attempt to address them: 

\textbf{Admission control:} In an open federation, MEC system providers may join or leave at any time. In centralized models, admission is regulated by the MFIP, which evaluates and approves new members. In decentralized models, admission is handled individually by each MEC. The main challenge in both cases is balancing the trade-off between open participation, security, and trust: strong admission controls add latency and overhead, while open policies increase the risk of spoofing and unauthorized access.

\textbf{Availability:} In dynamic environments, the number of federation participants changes frequently, making availability assurance difficult. Centralized models simplify participant monitoring but are vulnerable to outages due to their single point of failure. Decentralized models are more resilient to failures but complicate participant tracking and increase the risk of system integrity attacks.

\textbf{Security, privacy, and trust:} Each MEC provider manages its own security domain and often avoids disclosing detailed information about local infrastructure or full-service capabilities for competitive reasons. Centralized models can improve communication security via the MFIP, but require participants to share sensitive data, reducing privacy. Decentralized models preserve privacy by limiting data exchange, but results in lower security policies. Consistent mechanisms are needed to balance these concerns across domains.

\textbf{Billing management:} MEC providers aim to maximize profits by adjusting federation pricing based on demand, especially in dynamic environments. In centralized models, the MFIP can manage billing and act as an auctioneer, but this requires trust in and payment to a central authority. Decentralized models give pricing control to individual providers, but complicate dynamic SLA negotiation and secure billing.

\textbf{Multi-domain Quality of Service (QoS):} Ensuring QoS across federated MEC systems is particularly challenging in dynamic environments. Decentralized models often face difficulties in negotiating and enforcing SLAs on demand. The absence of a unified monitoring mechanism can lead to disputes that require third-party arbitration. Centralized models rely on a single authority to define, control, and monitor QoS across all participants.
\section{Applying Blockchain for MEC federation}
\label{sec:blockchain-for-mec-federation}
Depending on what is shared across domains, we distinguish between \emph{service federation} and \emph{resource federation}. In service federation, a consumer MEC system requests a service extension from a provider MEC system (e.g., application deployment). The provider deploys the requested service and establishes connectivity for the consumer to access and use the federated service. Once deployed, the provider acts as a proxy for the consumer and manages the service lifecycle. In contrast, resource federation involves sharing edge infrastructure resources (e.g., computing, networking), where the provider grants access and the consumer manages the deployment. This work focus on service federation rather than resource federation.

The process of service federation typically involves four key steps: \emph{(i)} MEC system registration, \emph{(ii)} service advertisement and discovery or announcement and negotiation, \emph{(iii)} service deployment, and \emph{(iv)} life-cycle management and charging.~\cite{dlt-federation-dynamic}. We further analyze steps \emph{(ii)} and \emph{(iii)} in Sec.~\ref{subsec:experimental-validation}, where we evaluate the performance of our solution in a scenario representing a service extension across different MEC systems through federation.

\begin{figure*}[t]
   \centering    
   \includegraphics[width=0.96\textwidth]{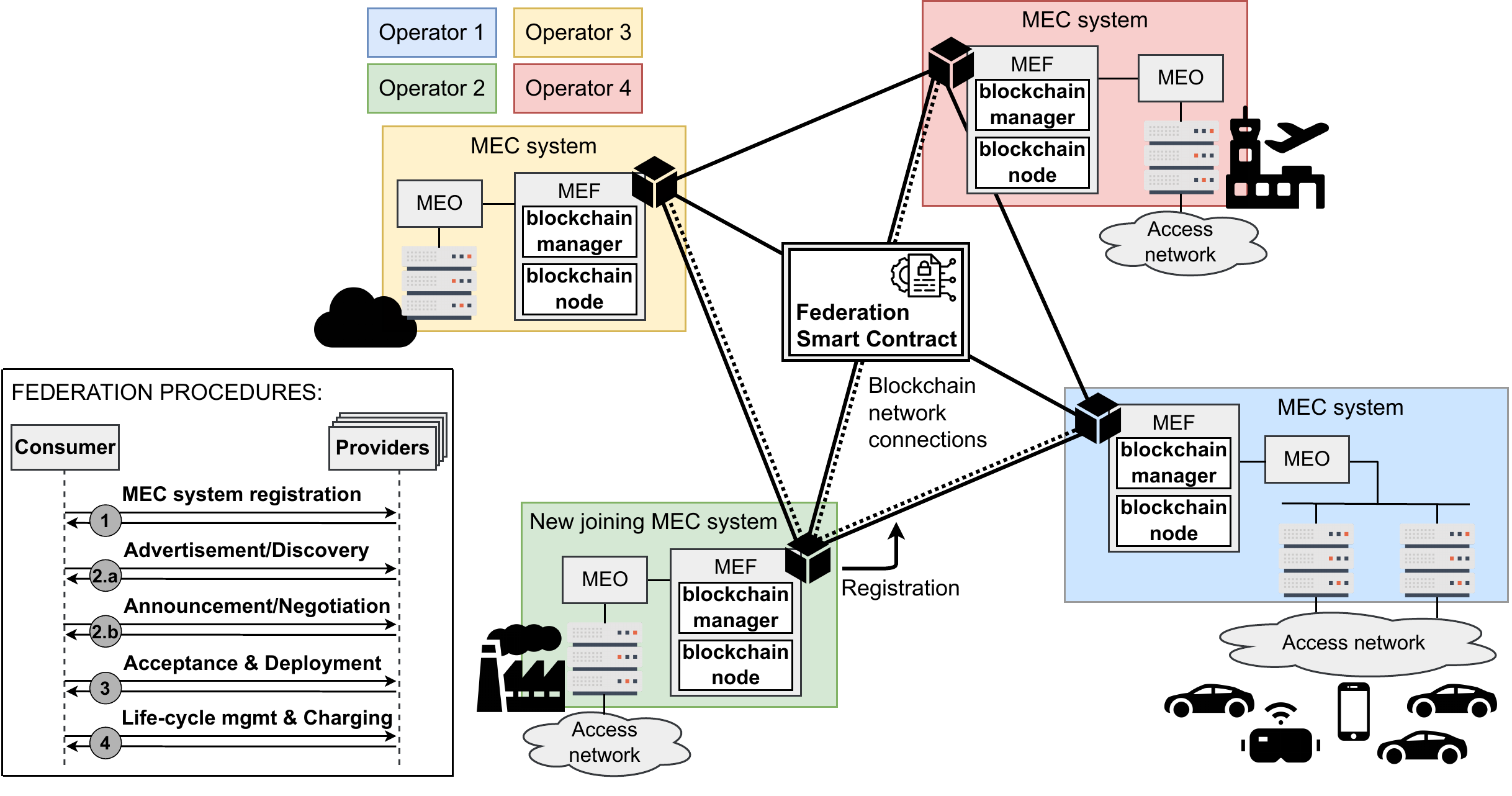}
   \caption{Blockchain architecture supporting MEC federation}
    \label{fig:proposed-solution}
\end{figure*}

\subsection{Blockchain overview}
\label{subsec:blockchain-overview}
Blockchain is a replicated database and computational platform shared among participants in a peer-to-peer network. Originally developed for Bitcoin, it provides a distributed, secure, and timestamped ledger that records transactions among anonymous users. Each block stores transaction data and a cryptographic reference to the previous block, forming an immutable chain back to the genesis block (block 0). 

There are two main types of blockchain networks: \emph{(i)} \textit{Permissionless}, where anyone can read, write, and participate in the creation of the ledger without prior approval; and \emph{(ii)} \textit{Permissioned}, where access and participation are restricted to trusted entities through identity and access controls. In the context of enterprise-grade mobile network infrastructures, permissioned chains are more suitable to ensure secure and controlled operation. 

All participants maintain a copy of the blockchain and synchronize updates using a consensus algorithm, which ensures global consistency across the network. In summary, the main benefits of using blockchain in networking include enhanced security, data integrity, transparency, elimination of intermediaries, and the integration of SCs~\cite{advantages-blockchain-networking}.

\subsection{The proposed blockchain solution for MEC federation}
\label{subsec:proposed-solution}
Most challenges associated with open and dynamic MEC federation (Sec.~\ref{subsec:mec-federation-challenges}) can be addressed by deploying a permissioned blockchain (e.g., Hyperledger, Enterprise Ethereum, Corda):

\textbf{Admission control} in the federation relies on the consensus and governance policies of the underlying blockchain. In permissioned chains, new members are admitted through consensus among existing participants. Although some nodes may act maliciously, they generally have an incentive to increase participation. If a consensus cannot be achieved, the federation may split into separate blockchain networks. 

\textbf{Availability} is guaranteed by operational and economic incentives for MEC systems to maintain active blockchain nodes, as it enables uninterrupted access to federated services and helps avoid financial or reputational penalties caused by downtime. Active participation also enhances network security by preventing centralization risks (e.g., 51\% attacks) and reduces operational costs (e.g., Ethereum gas fees) by minimizing transaction failures. In case of node failure, departure, or compromise, the blockchain network remains functional, allowing any MEC system to maintain connectivity via other nodes using its unique blockchain address.

\textbf{Security, privacy, and trust} are ensured through cryptographic mechanisms and usage limits. New MEC system providers face economic constraints (e.g., transaction fees, stake requirements) that limit their ability to send frequent transactions or federation announcements, helping to prevent spoofing and spam. Federation interactions are securely recorded as immutable transactions on the ledger, ensuring privacy and verifiable trust for all participants.

\textbf{Billing management and multi-domain QoS} can be realized through the use of SCs, which enable dynamic SLA definitions and automated QoS monitoring. MEC system providers, acting as contract owners, can dynamically adjust service pricing before customers initiate deposit transactions. To enforce inter-provider agreements, decentralized oracles -- networks of independent nodes that securely feed real-world data (e.g., SLA measurements from MEC deployments) to on-chain contracts -- can operate within the same ecosystem, as recommended by the ETSI’s Permissioned Distributed Ledgers group~\cite{advantages-blockchain-networking}. During service runtime, the SCs periodically queries the oracle network for updated QoS metrics. The oracle nodes collect and aggregate measurements from federated services off-chain, then return verified results on-chain. Based on this input, the contract evaluates SLA compliance and manages the payment process. In case of an SLA violation, the contract automatically issues a payment transaction to refund the penalty to the consumer's blockchain address.

The proposed system architecture for MEC federation is illustrated in Fig.~\ref{fig:proposed-solution}. It integrates a permissioned blockchain where each MEC system participates through an extended MEF that includes two additional components:

\begin{itemize}
    \item \textbf{Blockchain manager}: Acts as a bridge between the MEO and the blockchain layer. It translates federation operations into transactions and monitors blockchain events to notify the MEO of relevant updates.

    \item \textbf{Blockchain node}: Connects the MEC system to the network, participates in consensus, and maintains a synchronized local copy of the distributed ledger.
\end{itemize}

Federation procedures are implemented in a generic Federation SC deployed on the blockchain, acting as a distributed authority. The SC autonomously enforces federation logic and represents dynamic SLAs between provider and consumer MECs. All participating systems running a blockchain node execute the same instance of the SC simultaneously, ensuring global consistency, security, and trust. Please note that the SC design is fully transparent and crucial for safeguarding the privacy of sensitive information while overseeing federation procedures for all participants. 
\section{Proof-of-Concept Evaluation}
\label{sec:experimental-validation}
We implement our solution on top of Hyperledger Besu\footnote{\url{https://besu.hyperledger.org}}, an open-source framework for developing permissioned Ethereum blockchains, and evaluate its performance through a Proof-of-Concept (PoC) inspired by the Vehicle-to-Everything communication and Augmented Reality gaming use cases from ETSI GR MEC 035~\cite{etsi-gr-mec-035}. These scenarios highlight the need for service federation to enable direct communication between applications deployed across MEC systems from different mobile network operators. Unlike the public Ethereum Mainnet, Besu supports access-controlled deployments and enterprise-grade features that enable secure, high-performance transaction processing in a private network.

\subsection{Experimental setup and Workflow}
\label{subsec:experimental-validation}
\begin{figure}[tb]
   \centering \includegraphics[width=1\columnwidth]{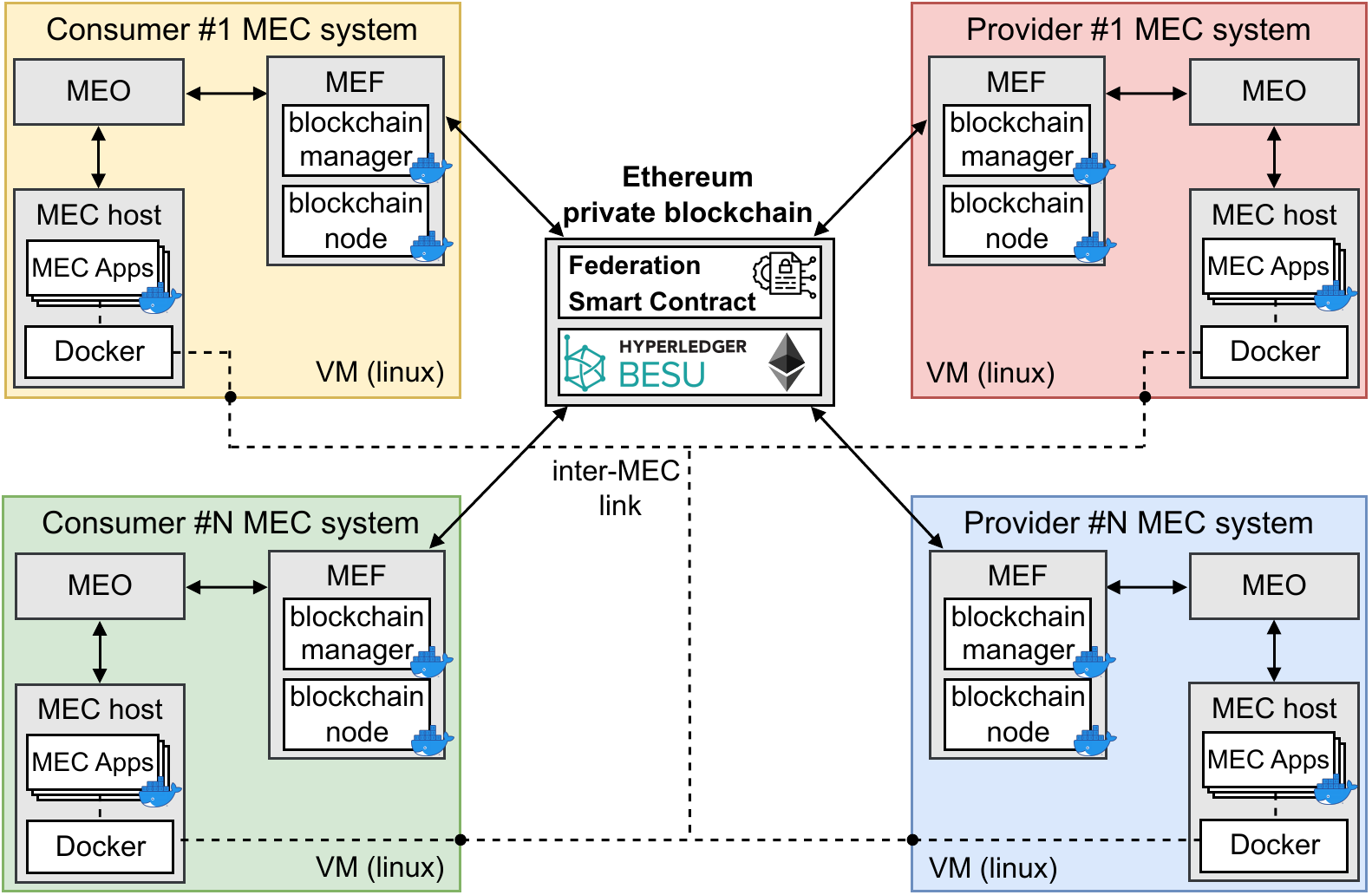}
   \caption{Experimental setup}
    \label{fig:experimental-setup}
    \vspace{-4mm}
\end{figure}

The PoC architecture (shown in Fig.~\ref{fig:experimental-setup}) consists of multiple MEC systems, each deployed as a Virtual Machine (VM) with 2~vCPUs, 2~GB RAM, and 20~GB storage. These VMs run on two physical servers equipped with Intel Xeon E5-2620~v4 CPUs (32~cores @~2.1GHz) and 64~GB RAM.

Each system includes: \emph{(i)} a MEC host using Docker as the VIM; \emph{(ii)} a custom orchestrator serving as the MEO to manage application lifecycle operations (creation, deletion, and modification) and the underlying infrastructure; and \emph{(iii)} a MEF composed of a blockchain manager and a Besu-based blockchain node to support federation, both deployed as Docker containers. Federated MEC services run as containerized applications using the Alpine Linux image, and the inter-MEC link for app-to-app data plane communication is established through an overlay network using the Virtual eXtensible Local Area Network (VXLAN) encapsulation protocol.

The MEF follows a cloud-native design decoupled from the MEO implementation, ensuring modularity and portability. The blockchain manager, built with the FastAPI\footnote{\url{https://fastapi.tiangolo.com}} web framework, provides the interface between the MEO and the blockchain layer. It exposes RESTful APIs for federation operations and uses the web3.py\footnote{\url{https://web3py.readthedocs.io/en/stable}} library to connect to the blockchain node via the Ethereum JSON-RPC API over WebSocket, supporting real-time subscription to SC events, state queries, and transaction submission. The blockchain node provides access to the blockchain, participates in the consensus, and ensures ledger synchronization.

At the consensus layer, we configured the Besu network with two popular algorithms: Clique and QBFT, which rely on a fixed validator set to propose and validate blocks. Both were configured with a block period of 5~s to balance latency and throughput while ensuring efficient transaction propagation as the network scales. The block period defines the time interval between the creation of two blocks and has a significant impact on overall blockchain performance. While shorter block periods can improve latency, setting them too low may lead to underfilled blocks or increased network overhead. To limit the impact of this trade-off on our results, we use this fixed setting in all experiments.

Note that the consensus is the primary factor influencing the speed and security characteristics of the blockchain. Clique is a lightweight algorithm where validators propose blocks in a round-robin order with minimal communication, but it does not guarantee immediate finality, allowing forks during network issues or validator downtime; whereas QBFT ensures stronger security through Byzantine fault tolerance, but relies on a multi-phase agreement process (proposal, pre-vote, and pre-commit) that introduces higher communication overhead. Based on these properties and our findings, we recommend protocols like Clique for trusted environments that prioritize rapid convergence toward federation agreements, and QBFT for low-trust scenarios requiring higher security guarantees. While in-depth consensus comparison is beyond the scope of this work, further discussion of consensus trade-offs in service federation can be found in~\cite{dlt-federation-consensus-comparison}.

\begin{figure}[tb]
   \centering \includegraphics[width=1\columnwidth]{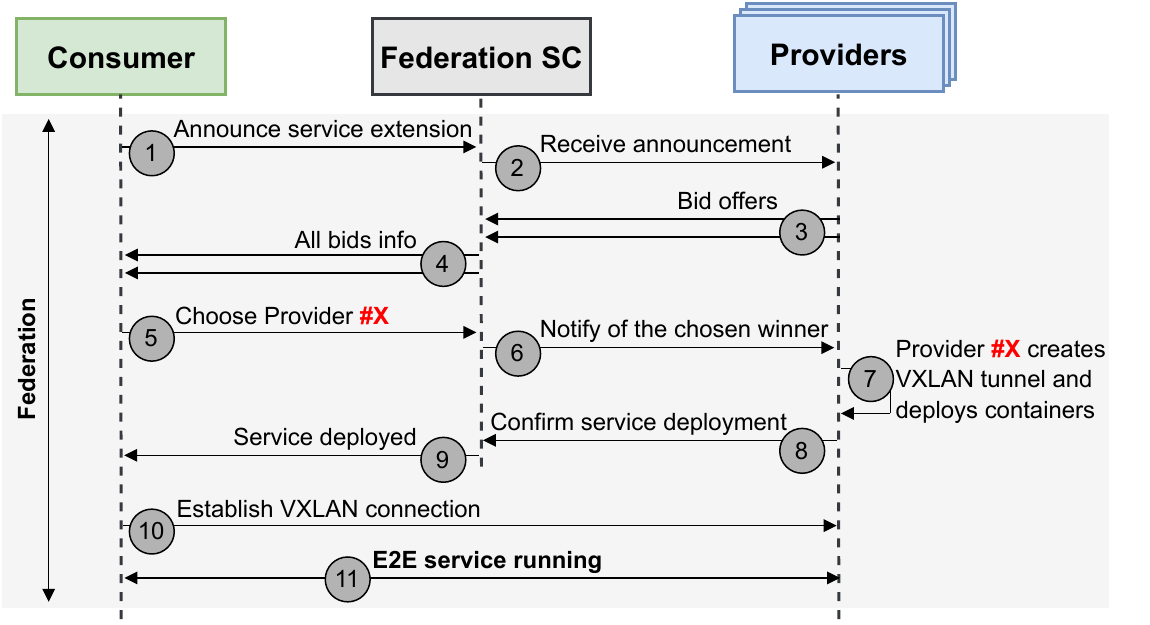}
   \caption{Workflow of Experimental setup}
    \label{fig:workflow-experimental-setup}
    \vspace{-4mm}
\end{figure}

Fig.~\ref{fig:workflow-experimental-setup} illustrates the execution workflow. Initially, the Besu network is active, the Federation SC is deployed, participants are registered with unique blockchain addresses, and consumer MECs host multiple applications. The federation starts when a consumer MEC (orchestrator) submits a transaction to the SC announcing the desired service extension (Step 1). Provider MECs subscribed to federation events receive the announcement (Step 2) and respond by submitting bid transactions with their service prices in a reverse-auction format (Step 3). 

To emulate real market dynamics, providers are modeled in different countries and submit bids that vary by location and time of day. Each offer is derived from the local electricity tariff adjusted by a time-of-day factor, introducing heterogeneous prices across providers and experiments. Once at least two offers are received (Step 4), the consumer selects the lowest-priced bid (Step 5).

The selected provider deploys the requested service (Steps 6–7) by creating a Docker bridge network, configuring its VXLAN tunnel with the consumer MEC using the information exchanged via the SC, and deploying the application containers onto the overlay network. After deployment, it sends a confirmation transaction (Step 8). The consumer receives it (Step 9), completes the VXLAN interconnection (Step 10), attaches the initial containers to the overlay network, and starts connection monitoring (Step 11). Federation is considered successful when continuous app-to-app communication is achieved between the consumer and provider MECs.

\subsection{Experimental Methodology}
\label{subsec:experimental-methodology}
We evaluate the performance of our framework in terms of latency, resource consumption, and scalability under increasing system load.

We measure latency as the total time required for a consumer to negotiate and execute a service federation (workflow in Fig.~\ref{fig:workflow-experimental-setup}). To assess the impact of the blockchain layer, we compare two popular consensus algorithms (Clique and QBFT) and quantify the latency overhead against a baseline MEF, dubbed hereafter State-of-the-Art (SOA), which removes all blockchain components while keeping the same experimental setup. In SOA, the MEF implements a federation manager that exposes REST APIs with other systems aligned with standard specifications and representative academic implementations~\cite{soa-enhanced-meo, soa-service-continuity-border-mobility, soa-mec-federation-resource-sharing}. All systems are mutually known in advance, with credential pairs stored in local registries for authentication. During federation, the consumer queries all registered peers for pricing models and requests service deployment from the lowest-cost provider via the federation interface. The provider deploys the service through its MEO, returns an HTTP confirmation, and the consumer establishes MEC app-to-app communication. For both MEF variants (blockchain and SOA), we execute 20 consecutive runs, logging on-chain transaction and event times (blockchain) and HTTP response times (SOA).

We consider scenarios with concurrent conditions where multiple consumers simultaneously issue service requests, increasing blockchain read/write operations and resulting in parallel federation executions. To emulate a realistic federation topology, we use the CAIDA AS Relationships dataset~\cite{caida-dataset}, which provides monthly snapshots since 1998 of business relationships between Autonomous Systems (ASes), i.e., networks managed by a single administrative domain that exchange routes via BGP. The dataset indicates that over $80\%$ of ASes are stubs (i.e., non-transit providers). Accordingly, we adopt an $80{:}20$ consumer-to-provider ratio and evaluate scenarios with MEC system count $N \in \{2, 10, 15, 25, 30\}$, corresponding to consumer--provider splits of (1,1), (8,2), (12,3), (20,5), and (24,6), respectively. We consider $N{=}30$ a reasonable upper bound, as the dataset reports that each AS connects to $\approx 8$ others on average.

Additionally, we profile the resource consumption of the blockchain node container on each system. For every run, we start CPU and memory monitoring on all nodes before triggering the workflow execution and stop it upon completion, sampling metrics every second via the Docker Engine API statistics stream\footnote{\url{https://docs.docker.com/reference/api/engine/version/v1.51/#tag/Container/operation/ContainerStats}}. CPU consumption is derived from Docker’s CPU deltas (i.e., the increase in container CPU time between consecutive samples divided by the increase in host CPU time) normalized by the number of online host cores and expressed in vCPUs. Memory consumption corresponds to the container’s active working set (excluding cached pages), reported in MB.

\begin{figure}[tb]
   \centering \includegraphics[width=1\columnwidth]{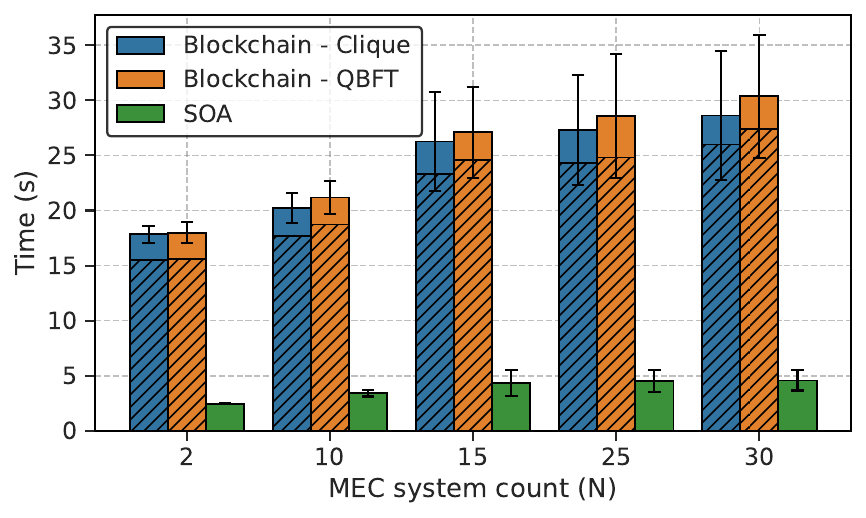}
   \caption{Federation time for different MEC counts, comparing blockchain MEF (with Clique and QBFT consensus algorithms) and baseline MEF (SOA). The results quantify the time overhead required for the fully automated negotiation and setup enabled by the blockchain approach.}
    \label{fig:comparison-federation-time}
    \vspace{-4mm}
\end{figure}

\subsection{Results Discussion}
\label{subsec:results}
Fig.~\ref{fig:comparison-federation-time} shows the average federation time (with variance) under different number of federation participants, comparing the blockchain MEF (using Clique and QBFT consensus) and the baseline MEF (SOA). A key insight from this figure is the composition of the blockchain bars. The hatched segments represent the blockchain-specific steps integral to our automated federation process: bidding, winner selection, deployment information exchange, and confirmation notification. These steps are not considered in typical state-of-the-art evaluations, which often assume the existence of pre-established, manually negotiated p2p contracts with defined prices. Therefore, the complete automation of the federation lifecycle in our solution introduces a quantifiable overhead, which ranges from a minimum of approximately 15.4 s (at N=2) to a maximum of 25.8 s (at N=30) compared to the SOA baseline. Conversely, the non-hatched segment of the bars represents the core instruction for service deployment. It is crucial to note that this portion of the process is directly comparable in duration to the total time of the SOA baseline. This demonstrates that once the automated negotiation is complete, our blockchain-based solution introduces no significant overhead to the actual service deployment and configuration tasks. Regarding the overall performance trends, the blockchain MEF with QBFT is consistently slower than Clique by less than 2 s, reflecting the coordination overhead required for QBFT’s multi-round consensus protocol. Furthermore, the federation time increases significantly up to N=15 before the growth becomes less pronounced. This is caused by initial load imbalance, where consumers overload a few low-cost providers, creating a queue for sequential operations like container start-up and VXLAN configuration. As more providers join, the load is distributed more evenly, and federation times stabilize. Finally, it is important to contextualize this federation overhead. The measured time represents the cost for achieving a dynamic and automated federation. This process does not impact the service's operational performance or continuity. The service downtime experienced by the end-user is strictly limited to the service migration window itself.

\begin{figure}[tb]
  \centering
  \subfloat[CPU consumption.]{%
    \includegraphics[width=0.49\linewidth]{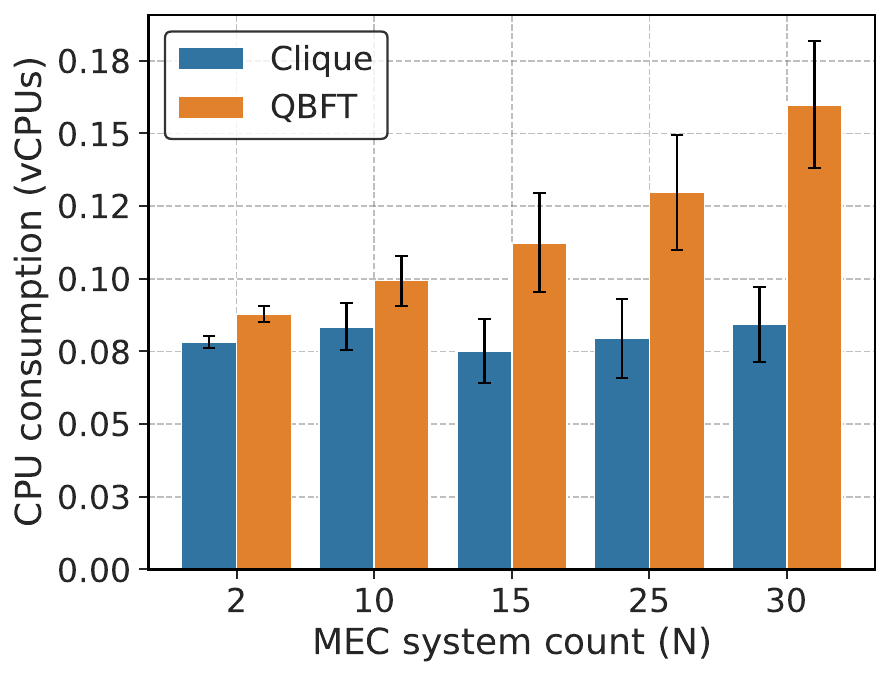}%
    \label{fig:cpu-consumption-per-blockchain-node-subfig}}
  \hfill
  \subfloat[Memory consumption.]{%
    \includegraphics[width=0.49\linewidth]{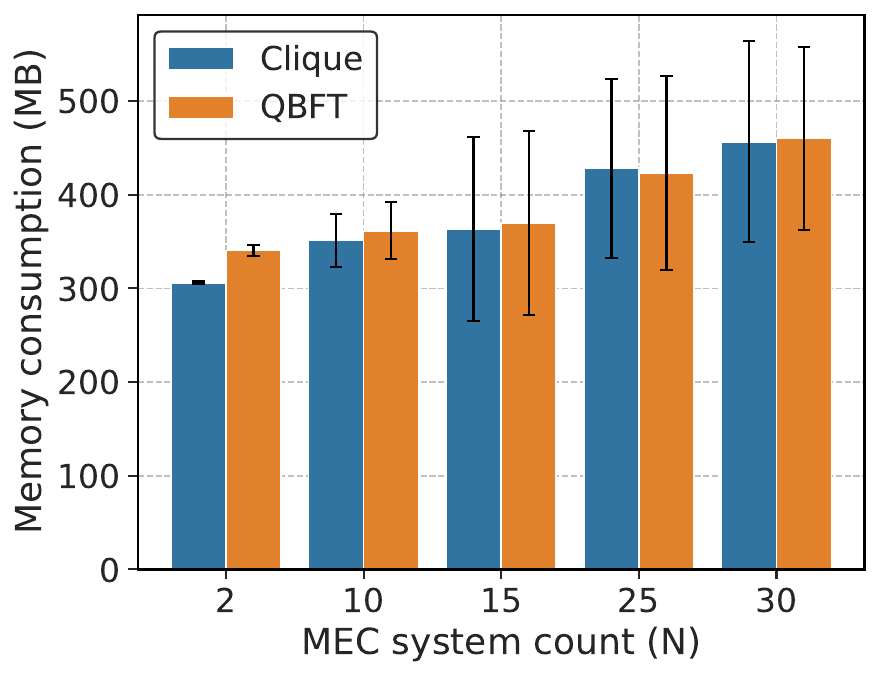}%
    \label{fig:ram-consumption-per-blockchain-node-subfig}}
  \caption{CPU and memory consumption per blockchain node during the federation workflow for different MEC counts and consensus algorithms.}
  \label{fig:resource-consumption-per-blockchain-node}
  \vspace{-2mm}
\end{figure}

Fig.~\ref{fig:resource-consumption-per-blockchain-node} presents the average CPU and memory consumption per blockchain node container during the federation workflow under different MEC system counts ($N$) and consensus algorithms (Clique and QBFT). In Fig.~\ref{fig:cpu-consumption-per-blockchain-node-subfig}, QBFT consistently consumes more CPU per node than Clique. Its overhead increases almost linearly with $N$, reaching $\approx0.16$~vCPUs at $N{=}30$, while Clique remains near $\approx0.08$~vCPUs with minor fluctuations. In contrast, Fig.~\ref{fig:ram-consumption-per-blockchain-node-subfig} shows that memory is the more critical resource, rising from 300~MB at $N{=}2$ to over 450~MB at $N{=}30$. This increase is primarily due to larger peer tables, message buffers, and ledger synchronization in bigger networks. The effect is amplified during concurrent federation requests, where larger blocks and transaction pools contribute further to memory usage.


\section{Related Work}
\label{sec:comparison-with-existing-work}

The concept of MEC federation has been explored in the literature. For instance,~\cite{soa-cooperative-service-continuity} relies on direct interfaces between MEC platforms, which introduce deployment and scalability challenges as the number of participants increases. Other works~\cite{soa-enhanced-meo, soa-cooperative-service-continuity, soa-mec-federation-resource-sharing} have proposed federation mechanisms aligned with standard specifications, including APIs and orchestrators. A common limitation, however, is their dependence on pre-established SLAs between MEC providers, making them suitable only for static environments. In contrast, our work targets open federation models and proposes a blockchain-based solution that enables MEC providers to quickly establish and break federation relationships without prior knowledge of each other. Centralized marketplaces have also been proposed for open federation~\cite{soa-market-auction-based}, where auctions coordinate service and resource matching. While promising, its centralized auction process risks single-point failures and complicates tenant auditability. Our approach instead uses SC to maintain immutable auction records, ensuring security and transparency.

Blockchain has been further studied for multi-domain network environments. The authors in~\cite{soa-blockchain-federation-1} propose a blockchain-based orchestration model for cross-domain service deployment in a scenario similar to ours, but evaluate only service instantiation. Unlike our work, they do not assess discovery, announcement, negotiation, or acceptance steps. The work in~\cite{soa-blockchain-federation-2} proposes a marketplace operated by a broker authority that, upon receiving requests from vertical users, deploys a dedicated SC to execute a reverse auction process. Infrastructure providers submit bids off-chain to the broker, which then records them on-chain and selects the winner, thereby granting the broker full and centralized control over the auction. Their implementation on a public Ethereum network also leads to high operational costs. Our approach distributes authority among providers and manages multiple auctions with a single persistent contract, removing reliance on a central broker. Following this direction,~\cite{soa-blockchain-federation-3} replaces the cenrtal broker with a consortium of network operators on a permissioned blockchain, further reducing costs. Nevertheless, their validation is limited to a simulation study and does not consider the impact of blockchain on the performance of the system.

In our previous work~\cite{dlt-federation-dynamic}, we studied the use of blockchain for service federation in dynamic NFV MANO (Network Function Virtualization Management and Orchestration) environments, focusing on the design and validation of the technology. We extended this in~\cite{dlt-federation-consensus-comparison} by evaluating the impact of different consensus algorithms on the same federation scenario. In this article, we shift focus to the ETSI MEC framework, introducing an enhanced blockchain-enabled MEF for dynamic federation, and evaluating its performance in terms of latency, resource consumption, and scalability.
\section{Conclusions}
\label{sec:conclusion}
This article presented a blockchain-based solution for edge federation that integrates smart contracts to enable private, secure, and efficient interactions among edge system providers. We validated the proposed design through a Proof-of-Concept implementation using an ETSI MEC-based environment and demonstrated its feasibility and scalability. Future work will involve real-world deployment of service federation and performance comparison with alternative blockchain platforms (e.g., Hyperledger Fabric).

\section*{Acknowledgements}

\bibliographystyle{IEEEtran} 
\bibliography{references} 

\begin{thebibliography}{10}
\providecommand{\url}[1]{#1}
\csname url@samestyle\endcsname
\providecommand{\newblock}{\relax}
\providecommand{\bibinfo}[2]{#2}
\providecommand{\BIBentrySTDinterwordspacing}{\spaceskip=0pt\relax}
\providecommand{\BIBentryALTinterwordstretchfactor}{4}
\providecommand{\BIBentryALTinterwordspacing}{\spaceskip=\fontdimen2\font plus
\BIBentryALTinterwordstretchfactor\fontdimen3\font minus \fontdimen4\font\relax}
\providecommand{\BIBforeignlanguage}[2]{{%
\expandafter\ifx\csname l@#1\endcsname\relax
\typeout{** WARNING: IEEEtran.bst: No hyphenation pattern has been}%
\typeout{** loaded for the language `#1'. Using the pattern for}%
\typeout{** the default language instead.}%
\else
\language=\csname l@#1\endcsname
\fi
#2}}
\providecommand{\BIBdecl}{\relax}
\BIBdecl

\bibitem{etsi-gr-mec-035}
\BIBentryALTinterwordspacing
{ETSI}, ``{Multi-access Edge Computing (MEC); Study on Inter-MEC systems and MEC-Cloud systems coordination},'' {ETSI}, Group Rep. GR MEC 035 V3.1.1, 2021, {Accessed: Aug. 27, 2025}. [Online]. Available: \url{https://www.etsi.org/deliver/etsi_gr/MEC/001_099/035/03.01.01_60/gr_mec035v030101p.pdf}
\BIBentrySTDinterwordspacing

\bibitem{gsma-federation}
\BIBentryALTinterwordspacing
{GSMA}, ``{Telco Edge Cloud: Edge Service Description and Commercial Principles},'' {GSMA}, White Paper, 2020, {Accessed: Aug. 27, 2025}. [Online]. Available: \url{https://www.gsma.com/solutions-and-impact/technologies/networks/wp-content/uploads/2020/10/GSMA-Telco-Edge-Service-Description-Commercial-Principles-Oct-2020.pdf}
\BIBentrySTDinterwordspacing

\bibitem{soa-service-continuity-border-mobility}
K.~Rasol \emph{et~al.}, ``Mec federation for seamless service continuity in cross-border mobility scenarios,'' in \emph{IEEE Future Networks World Forum (FNWF)}.\hskip 1em plus 0.5em minus 0.4em\relax IEEE, 2024, pp. 675--680.

\bibitem{soa-enhanced-meo}
J.~Palomares, E.~Coronado, C.~Cervell{\'o}-Pastor, E.~Carmona-Cejudo, and S.~Siddiqui, ``{MEO: An Enhanced MEC Orchestrator for Federated and Distributed MEC Systems},'' in \emph{IEEE Global Communications Conference}.\hskip 1em plus 0.5em minus 0.4em\relax IEEE, 2024, pp. 5325--5330.

\bibitem{soa-cooperative-service-continuity}
S.~M. Hosseini \emph{et~al.}, ``{Cooperative, Connected and Automated Mobility Service Continuity in a Cross-Border Multi-Access Edge Computing Federation Scenario},'' \emph{Frontiers in Future Transportation}, vol.~3, p. 911923, 2022.

\bibitem{soa-mec-federation-resource-sharing}
R.~Rodriguez, A.~Cl{\'e}rigo, P.~Rito, S.~Sargento, B.~Parreira, and R.~Dinis, ``{MEC Federation: A Framework for Resource Sharing in Multi-Operator Beyond-5G Networks},'' in \emph{NOMS 2025-2025 IEEE Network Operations and Management Symposium}.\hskip 1em plus 0.5em minus 0.4em\relax IEEE, 2025, pp. 1--7.

\bibitem{etsi-gs-mec-003}
\BIBentryALTinterwordspacing
{ETSI}, ``{Multi-access Edge Computing (MEC); Framework and Reference Architecture},'' {ETSI}, Group Spec. GS MEC 003 V3.1.1, 2022, {Accessed: Aug. 27, 2025}. [Online]. Available: \url{https://www.etsi.org/deliver/etsi_gs/MEC/001_099/003/03.01.01_60/gs_mec003v030101p.pdf}
\BIBentrySTDinterwordspacing

\bibitem{dlt-federation-dynamic}
K.~Antevski and C.~J. Bernardos, ``{Federation in dynamic environments: Can blockchain be the solution?}'' \emph{IEEE Communications Magazine}, vol.~60, pp. 32--38, 2022.

\bibitem{advantages-blockchain-networking}
{ETSI}, ``{Permissioned Distributed Ledgers (PDL); Smart Contracts System Architecture and Functional Specification},'' ETSI, Group Rep. {GR PDL 004 V1.1.1}, 2021.

\bibitem{dlt-federation-consensus-comparison}
K.~Antevski and C.~J. Bernardos, ``{Applying Blockchain consensus mechanisms to Network Service Federation: Analysis and performance evaluation},'' \emph{Computer networks}, vol. 234, p. 109913, 2023.

\bibitem{caida-dataset}
``{The CAIDA AS Relationships Dataset},'' \url{https://www.caida.org/catalog/datasets/as-relationships}, 2024, {Accessed: Aug. 29, 2025}.

\bibitem{soa-market-auction-based}
M.~Dieye, W.~Jaafar, H.~Elbiaze, and R.~H. Glitho, ``{Market Driven Multidomain Network Service Orchestration in 5G Networks},'' \emph{IEEE Journal on Selected Areas in Communications}, vol.~38, pp. 1417--1431, 2020.

\bibitem{soa-blockchain-federation-1}
R.~V. Rosa and C.~E. Rothenberg, ``{Blockchain-Based Decentralized Applications for Multiple Administrative Domain Networking},'' \emph{IEEE Communications Standards Magazine}, vol.~2, pp. 29--37, 2018.

\bibitem{soa-blockchain-federation-2}
M.~F. Franco, E.~J. Scheid, L.~Z. Granville, and B.~Stiller, ``{BRAIN: Blockchain-based reverse auction for infrastructure supply in virtual network functions-as-a-service},'' in \emph{IFIP Networking Conference}.\hskip 1em plus 0.5em minus 0.4em\relax IEEE, 2019, pp. 1--9.

\bibitem{soa-blockchain-federation-3}
M.~A. Togou \emph{et~al.}, ``A distributed blockchain-based broker for efficient resource provisioning in 5g networks,'' in \emph{International wireless communications and mobile computing (IWCMC)}.\hskip 1em plus 0.5em minus 0.4em\relax IEEE, 2020, pp. 1485--1490.

\end{thebibliography}

\section*{Biographies}

\small{
  \textbf{Adam Zahir} (azahir@pa.uc3m.es) received his M.Sc. in 2024 and is a Ph.D. student at Universidad Carlos III de Madrid (UC3M).
  
  \textbf{Milan Groshev} received his M.Sc. in 2016 and Ph.D. in 2022, and is a postdoctoral researcher at IE University.
  
  \textbf{Carlos J. Bernardos} received his M.Sc. in 2003 and Ph.D. in 2006, and is an associate professor at UC3M.
  
  \textbf{Antonio de la Oliva} received his M.Sc. in 2004 and Ph.D. in 2008, and is an associate professor at UC3M.
}

\end{document}